\documentclass{bmcart}
\pdfoutput=1
\usepackage{xcolor}
\usepackage{amsthm,amsmath}
\usepackage{amsfonts}
\usepackage{moreverb,url}

\usepackage{natbib}

\usepackage[utf8]{inputenc} 
\usepackage{graphicx} 

\usepackage{parskip}
\begin{document}

\begin{frontmatter}

\begin{fmbox}
\dochead{Research}

\title{A bi-directional approach to comparing the modular structure of networks}

\author[
 addressref={CASA}, 
 noteref={n1}, 
 email={d.straulino@ucl.ac.uk} 
]{\inits{DS}\fnm{Daniel} \snm{Straulino}}
\author[
 addressref={Ox},
 email={mattie.landman@maths.ox.ac.uk}
]{\inits{ML}\fnm{Mattie} \snm{Landman}}
\author[
 addressref={CASA,Ox,ATI}, 
 email={oclery@maths.ox.ac.uk}
]{\inits{NO}\fnm{Neave} \snm{O'Clery}}

\address[id=CASA]{
 \orgname{Center of Advanced Spatial Analysis, University College London}, 
 \street{90 Tottencourt Road, Bloomsbury}, %
 \postcode{W1T 4TJ} 
 \city{London}, 
 \cny{UK} 
}

\address[id=Ox]{
 \orgname{Mathematical Institute, Oxford University}, 
 \street{Andrew Wiles Building, Radcliffe Observatory Quarter}, %
 \postcode{OX2 6GG} 
 \city{Oxford}, 
 \cny{UK} 
}

\address[id=ATI]{
 \orgname{Alan Turing Institute}, 
 \street{British Library, 96 Euston Road}, %
 \postcode{NW1 2DB} 
 \city{London}, 
 \cny{UK} 
}

\begin{artnotes}
\note[id=n1]{Corresponding author} 
\end{artnotes}
\end{fmbox}

\begin{abstractbox}
\begin{abstract} 

Here we propose a new method to compare the modular structure of a pair of node-aligned networks. The majority of current methods, such as normalized mutual information, compare two node partitions derived from a community detection algorithm yet ignore the respective underlying network topologies. Addressing this gap, our method deploys a community detection quality function to assess the fit of each node partition with respect to the other network's connectivity structure. Specifically, for two networks A and B, we project the node partition of B onto the connectivity structure of A. By evaluating the fit of B's partition relative to A's own partition on network A (using a standard quality function), we quantify how well network A describes the modular structure of B. Repeating this in the other direction, we obtain a two-dimensional distance measure, the bi-directional (BiDir) distance. 
The advantages of our methodology are three-fold. First, it is adaptable to a wide class of community detection algorithms that seek to optimize an objective function. Second, it takes into account the network structure, specifically the strength of the connections within and between communities, and can thus capture differences between networks with similar partitions but where one of them might have a more defined or robust community structure. Third, it can also identify cases in which dissimilar optimal partitions hide the fact that the underlying community structure of both networks is relatively similar.
We illustrate our method for a variety of community detection algorithms, including multi-resolution approaches, and a range of both simulated and real world networks.

\end{abstract}


\end{abstractbox}

\end{frontmatter}

\section*{Introduction}

The last thirty years have been extremely fruitful in the study of networks, which enable us to represent and analyse the complex interconnection structure of a wide range of real world and engineered systems \cite{newman2006structure,strogatz2001exploring,newman2003structure}. Such is the versatility and usefulness of such a representation, network analysis is increasingly popular in fields as diverse as physics, biology and sociology (see e.g., \cite{barabasi1999mean,dunne2002network,jeong2000large,krause1997soziale,sood2005voter}). While the networks toolbox continues to expand, particularly with respect to dynamics-based approaches \cite{barahona2002synchronization,o2013observability} and statistical methods \cite{robins2007recent,cranmer2011inferential}, there remains a range of methodological challenges related to common applications. Seeking to fill an apparent gap, here we develop a new method to address the simple and widely applicable question of the comparison of the community structure of two networks with an identical set of nodes. 

A node partition into communities corresponds to a division of nodes into non-overlapping sets such that each set of nodes is tightly interconnected with few connections to the rest of the network \cite{girvan2002community}. The presence of such communities is typically linked to the presence of some kind of higher-order organisation in a network, and is often related to functional or societal structure \cite{palla2005uncovering}. In many cases, a network may be reduced to a community-based representation which retains the fundamental structural and dynamical features of the original network \cite{fortunato2010community,mucha2009communities}. There exist a wide variety of methods to detect community structure, as reviewed by Schaub \emph{et al} \cite{schaub2017many}. 

Here we focus on developing a new method to compare the modular structure of networks. This is key to, for example, understanding differences in brain function in two neural networks, or understanding how partisanship has changed over time in the United States Congress. Since the presence of communities in these networks lies at the core of the question of interest, it is natural to use them as the basis of comparison. Perhaps the most widely-used community comparison technique is \textit{normalized mutual information} (NMI) \citep{Danon2005}. This and other similar techniques are derived from an information theoretic approach to comparing sets. By focusing exclusively on partitions, and ignoring all other features of the network (for example, the strength of the connections between different communities), these methods lose information. Furthermore, while mutual information is symmetric, there are frequent cases where one network provides useful information about the other but not vice-versa. For example, if the communities of one network are nested inside the communities of the other (meaning that one community of the second network contains several communities of the first) then the second network provides information about the structure of the first but not the other way around.

We propose a methodology that addresses both of these issues, and that is agnostic to the use of community detection algorithm within a large class of commonly used approaches. We call it the \textit{bi-directional distance}, and it is based on a simple idea: we swap the partitions representing the communities of the two networks and evaluate whether the re-assigned communities are a good fit. In the sections below we demonstrate this method for a range of toy and synthetic networks, as well as a real-world example concerning inter-industry labour flows relevant for industrial policy. Labour flow networks illuminate the labour mobility structure of an economy, which is an important factor in the transport of industrial know-how and the localised emergence of new economic activities at the heart of regional economic growth policies. 

The paper is structured as follows. We begin by giving a short overview of both the field of community detection and current network comparison methodologies, while also illustrating the shortcomings of current community comparison techniques. We then present our new methodology and the results of experiments on toy and synthetic networks. Second, we demonstrate the flexibility and wide applicability of our method using three distinct community detection algorithms. Third, we apply our methodology to compare the inter-industry labour flow networks of 5 European countries, Germany, Sweden, the Netherlands, the UK and Ireland. Finally, we extend our methodology to the more complex case of multi-scale community detection.

\section*{Literature Review}

\subsection*{Community detection}

Given the diversity of community detection algorithms, it is useful to classify approaches according to the nature of the problem at hand. Schuab \emph{et al} \cite{schaub2017many} propose four broad categories of community detection objective: community detection as the minimization of some constraint function, community detection as clustering into densely connected groups, community detection aimed at identifying structurally similar nodes and community detection as the simplified description of dynamical flows on the network. We follow this categorization to briefly introduce some of the most popular community detection algorithms.

In the first category we find cut based approaches which aim to minimize the number or weight of the edges between communities. In other words, the constraint to minimize is calculated as the number of edges that must be deleted (or cut) to achieve the partition, e.g., ``ratio-cut" \citep{hagen1992new}. Graph partitioning is heavily used in parallel computing, in the layout of circuits, and to design serial algorithms to solve partial differential equations \cite{fortunato2010community}. 

In the second category we find, among others, modularity optimization \citep{newman2004finding} which is one of the most popular community detection algorithms. A key difference with cut based approaches is that modularity does not a priori set the number of clusters, nor constrains them to be of similar size. Although not without its limitations, this approach has seen widespread use and given rise to a large number of variations and adaptations \citep{fortunato2010community}. It has been   used to investigate protein interaction networks, functional brain networks, and ecological networks among other applications.  

The structural category groups nodes that exhibit similar connectivity patterns, that is, they are similar if they connect to similar nodes with equal probability . One popular technique for identifying groups in this manner is the stochastic block model (SBM) \citep{holland1983stochastic}. This approach is based on the assumption that nodes can be divided into certain classes, and that the likelihood of a connection between two nodes is determined by their respective classes. This gives rise to a probabilistic model for the network such that, via identification of the best fitting parameters, latent groups in the network can be recovered. Originating in the social networks literature, this approach is frequently used to uncover sub-groups in a social network.

The last category deploys dynamics on networks to uncover modular structure  \cite{newman2006structure,masuda2017random}. In this category we find Markov stability \citep{delvenne2010stability} and InfoMap \cite{Rosvall2009}. Markov stability, for example, exploits random walker dynamics on a network to detect the presence of node communities. If a walker, which jumps from node to node with probability proportional to edge weight, gets 'trapped' in a region for a period of time this indicates a cluster of nodes with high internal connectivity. This approach has been applied to dynamical processes on networks such as consensus and synchronisation (e.g., \cite{Barahona2003,oclery2016}). 

In general, as we have seen, different types of application call for different approaches to detecting community structure. Furthermore, no particular methodology consistently outperforms others \citep{peel2017ground}. Hence, any network comparison technique that focuses on the comparison of modular structure will need to accommodate a variety of underlying community detection techniques.

\subsection*{Network comparison methods}

There is a rich literature in network comparison \citep{donnat2018tracking} spanning from global methods (e.g., spectral distances) to more localised approaches (e.g., graph editing). Since we are interested in comparing labelled networks, where each node represents a particular instance (a person, a place, etc.), we focus on methods where there is a one to one correspondence between the nodes of both networks. 

Perhaps the simplest approach is to count the proportion of edges that need to be added and deleted to transform one graph into the other, a measure which originated in the field of information theory and is known as the Hamming distance \citep{hamming1950error}. While easy to interpret, the Hamming distance is highly sensitive to the density of the graphs, a shortfall that is addressed by the Jaccard distance \citep{jaccard1912distribution}. The simplicity of these methods has seen them become very popular, not just in the network comparison literature, but also in the fields of information theory and machine learning \citep{donnat2018tracking}. Nevertheless, they are very local in scope, looking only at the direct neighbours of each node, with all deletions and insertions given equal weight regardless of changes in properties such as node importance as captured by various centrality measures. Therefore, these methods are not well suited to capture similarities at medium or large scales.

Motif based methods, which compare the frequency of certain sub-graphs (motifs) such as triangles or stars, have proved very popular in applications to protein interaction networks \citep{milo2002network}. Despite being relatively local in scope, they are able to capture patterns beyond immediate neighbourhoods. This technique is particularly useful when the motifs represent functional components of the network. However, just as for graph editing distances, these methods fail to capture similarities in large scale connectivity or organisation patterns across networks.

On the other end of the spectrum we have spectral distances. These are global measures based on the eigenvalues of either the adjacency matrix or some version of its Laplacian. Since the eigenvalues are related to topological properties of the network, such as connectivity, they capture global features of the network. Nevertheless, these approaches are permutation invariant, and thus ignore the labels of the nodes which renders them impractical for many applications (consider a network of friendships, the spectral distance does not differentiate between the original network and one obtained by randomly exchanging individuals). 

\begin{figure}[h]
    \centering
    \includegraphics[width=1\linewidth]{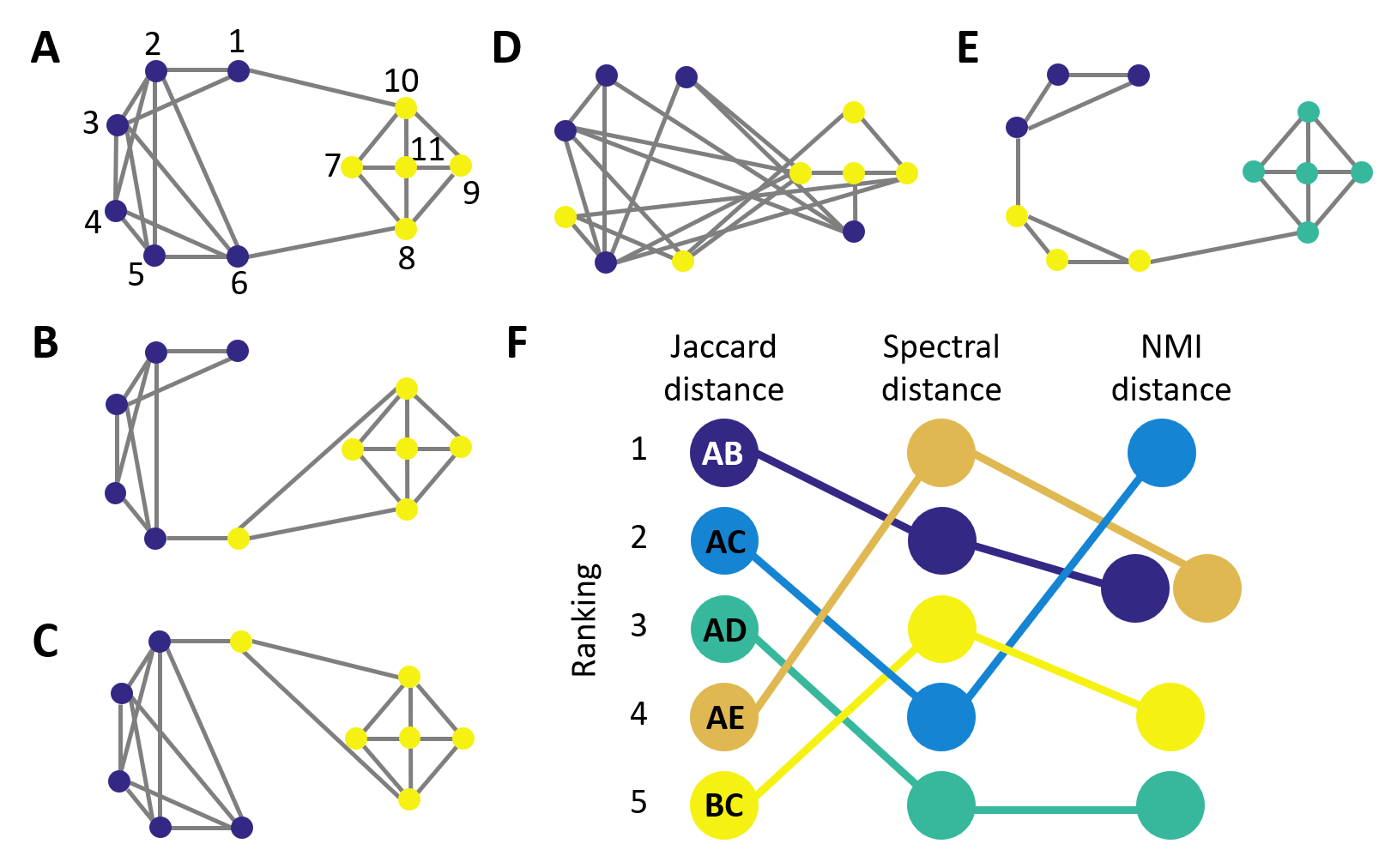}
    \caption{Here we show five networks whose nodes are coloured according to their communities found using modularity. We illustrate the different behaviours of three network comparison methodologies: the Jaccard distance (graph editing), the Spectral distance (eigenvalues) and the NMI distance (partitions). Sub-figure F shows the ranking of these similarity metrics for network pairs from closest (top) to furthest (bottom). Notice that standard community comparison metrics like NMI do not differentiate between pair A-B and pair A-C since they ignore the underlying structure and consider exclusively the partitions.}
    \label{ToyMotivation}
\end{figure}

There are other approaches, both global and local, beyond those that we have outlined above but there is comparatively little work comparing networks at a mesoscopic or modular scale. By far the most widely-used community-based comparison technique is \textit{Normalized Mutual Information} (NMI) \citep{Danon2005}. NMI, which has been borrowed from the field of information theory, defines the similarity of two partitions as the mutual information of the two partitions, normalized by a combination of the entropy of each partition. It therefore measures how easily one can infer one partition, given the other. Modifications of NMI have been proposed to address some of its limitations, namely that it can do very poorly when the number of communities in two networks differ \cite{onnela2012taxonomies,newman2020improved}. 

Figure \ref{ToyMotivation} illustrates some key network comparison methodologies. We have chosen three popular methodologies: the Jaccard distance, the Spectral distance, and NMI. While the Jaccard distance considers $B$ to be the most similar to $A$, the Laplacian ranks $C$ as more similar, and NMI indicates that $E$ is the closest network to $A$. It is clear that these metrics capture different features of the networks. 
We are interested in comparing the community structure, which is shown via the colouring of the nodes in the figure. We observe that $E$ is mostly embedded in $A$ and therefore quite similar. This is picked up by NMI. Comparing $B$ and $C$ to $A$, it appears that $C$ is 'closer' since the node that changes community is relatively peripheral to both the blue and yellow clusters. However, since NMI only uses information on partitions - and ignores the strength of community as well as the role of the nodes - it does not differentiate when comparing $A$ to $B$ or $C$.

Although the above methodologies for comparing networks based on community structure have proven to be useful, the fact remains that there are a number of shortcomings. Perhaps the most prominent issue is that NMI and its modifications rely exclusively on partitions of network, and ignore the quality of these partitions. By reducing the community structure to a partition, these methodologies discard important information about the network \cite{orman2012comparative}.  There have been some efforts to address this issue \cite{chakraborty2017metrics}, but they rely on correction factors (usually weighting the nodes by their degree). Furthermore, we described earlier how different community detection algorithms differ in their objectives, but partition based methodologies do not account for these differences. 

In this paper we propose an alternative framework to comparing networks: partition swapping. This approach presents three marked advantages: it is simple to understand and compute, it incorporates information beyond the partition of the network, and it is flexible, adapting to different community detection algorithms. Its simplicity comes from the underlying idea: for a pair of networks, we assess the fit of each node partition with respect to the other network's connectivity structure. Since the quality function takes into account the edge weights in order to evaluate the fit, we include information about the network topology. And by changing the quality function we immediately adapt to the application at hand. 

\subsection*{Inter-industry labour flow networks}

One of the initial motivations for the development of a new modular network comparison method was to compare the community structure of several industry networks from different countries. The edges of these industry networks are derived from worker data on inter-industry job switches. The extent to which workers jump between industry pairs corresponds to the degree of skill-overlap, a key metric for the 'relatedness' between industries which underlies network-based modelling of regional growth paths \cite{frenken2007,neffke2011regions}. This type of network was first introduced by Neffke \emph{et al} \cite{Neffke2013SkillRelatedness} and is known as a skill-relatedness network (SRN) in the literature. A key feature of these networks is their modular structure \cite{ocleryIreland}, which constrains regional diversification paths and knowledge diffusion. One might expect that these networks exhibit a near-universal structure, with skill-based industry clusters conserved across countries. On the other hand, differences in structure may illuminate potential growth paths and opportunities unseen in models based solely on data from a single country. Here we investigate these questions, seeking to deploy our new methodology to illuminate both similarities and differences across networks.   

Industry networks where nodes are connected based on some kind of skill or capacity overlap have emerged as a key modelling tool in evolutionary economic geography. The key idea is that regions are constrained by the capabilities and know-how embedded in their workforce and industrial structure \citep{NelsonWinter1982, frenken2007, hausmann2007}. In order to diversity and grow, regions tend to move into new industries that share similar or 'related' capabilities in an incremental path dependent fashion \cite{hidalgo2007product, Neffke2013SkillRelatedness, neffke2011regions}. In order to model this process, a range of different types of industry networks capturing the 'relatedness' between industry pairs have been proposed. These include, amongst others, geographic clustering or co-location of industries as a proxy for general capability overlap \citep{hausmann2007}, occupational similarity as a proxy for labour sharing \citep{farjoun1994beyond}, and collaboration on patents as a proxy for knowledge sharing \citep{jaffe1989characterizing}. Here we focus on 'skill-relatedness' which was proposed by Neffke \emph{et al} \citep{neffke2013skill}. Intuitively, if many workers move from one industry to another, then it is likely that these industries share a high degree of skill-similarity. This metric is relatively less noisy and has more even industry coverage relative to other alternatives. 

Network-based models for regional growth and diversification paths broadly work as follows. We can think of the 'location' of a region in an industry network in terms of the collection (subgraph) of industries it currently has. Potential new industries share existing skills and capabilities, and correspond to neighbouring nodes in the network. Hence, if a region has many industries on the periphery of the network, it has few related industries into which it is able diversify. However, if a region has industries which are more central and connected, it has many more potential diversification paths. Skill-relatedness networks have been applied to model and predict the diversification paths of regions in a large number of contexts \citep{Neffke2013SkillRelatedness, neffke2011, o2019modular}. They have also been used in various other applications, such as models for urban formality growth \citep{oclery2016,o2019commuting}, employment resilience and adaptability \citep{diodato2014resilience} and knowledge spillovers from MNEs \citep{csafordi2016effect}. 

Recently, O'Clery \emph{emph}{et al} \cite{o2019modular} showed that these networks have a distinct modular structure. Hence, they contain communities (i.e., groups of industries) within which workers switch much more easily (relative to switches between industries in different communities). The authors refer to these communities as `skills-basins' in reference to the degree of skill and knowledge sharing within industry groupings. The presence of modular structure is a key topological feature of these networks that constrains the industrial diversification opportunities of regions, and can be used to improve network-based models for regional diversification and growth patterns. 

In this study, we are interested in comparing the modular structure of various countries' SRNs. Within the related diversification literature, the SRN has been assumed to be identical across both space and time. Neffke \emph{et al.}\citep{neffke2011} showed that there was little variation between the SRN when constructed for East and West of Germany, as well as for Germany compared to Sweden \citep{Neffke2013SkillRelatedness}. The authors estimated the similarity between the two networks by calculating the Spearman correlation between edges. Based on these results, various studies have used the SRN of a different country within their analysis when the SRN of the country under consideration was not available. 

We believe, however, that these networks may be different, particularly on a mesoscopic-scale (i.e., with respect to their modular structure). As an SRN is constructed from all inter-industry labour flows within a country, the network reflects intricacies of the structure of the local labour market. We expect that the degree of inter-industry labour flows is highly dependent on the historical economic progress of a country and its institutional labour market structures. We are interested in both similarities and differences in the modular structure of these networks. Similarities uncover universal structure in inter-industry skill-sharing, and portability of networks across contexts. Differences may provide insight into hidden growth opportunities: linkages and clustering patterns present in one context may suggest potential unseen paths in another.

\section*{Results}

\subsection*{A bi-directional distance metric}

A large class of community detection algorithms are based on optimizing an objective function $F$ that measures the goodness of fit of a partition according to some desired property, whether structural (for example modularity from \cite{newman2006modularity}), dynamic (see \cite{delvenne2010stability}) or other (see \cite{rosvall2009map}). We propose to compare the modular structure of two networks, say A and B, by computing the ratio of A's $F$-score under B's optimal partition to its F-score under its own optimal partition, and vice versa. In a sense, our two dimensional method tells us how well network A describes modular structure of B as well as the other way around. 

More formally, let $A,B \in M^{n \times n}$ be the adjacency matrices of two networks of size $n$, where there is a one to one correspondence of the nodes in each network given by the identity function. In other words, the labelling of the nodes matter and we assume that the labels of both networks are identical. Let $\mathcal{P}$ be the space of partitions of $[n]$, the set of integers up to $n$. Without loss of generality, let $F:M^{n \times n}\times\mathcal{P} \mapsto[0,1]$ be an objective function. We then denote
\[
P_A := \underset{P \in \mathcal{P}}{\arg\max} \, F(A,P),
\]
\[
P_B := \underset{P \in \mathcal{P}}{\arg\max} \, F(B,P),
\]
the partitions that maximize the objective function for each network. We are now ready to define our (non-symmetric) distance score:
\begin{equation}
d(A,B) = 1-\frac{F(A,P_B)}{F(A,P_A)}.
\end{equation}
The ratio on the right tells us how well network A describes modular structure of B - a high value indicates that the optimal partition of B is indeed also a good partition of network A. By construction, $d(A,B)$ ranges over $[0,1]$, and is $0$ if and only if $F(A,P_A)=F(A,P_B)$. By swapping $A$ and $B$ we can obtain a second distance score. In this way, for every pair of networks, we have a pair of distance scores that reflect how well their respective partitions capture each others' community structure. We propose a two dimensional distance, the partition swapping score \textit{BiDir}:
\begin{equation}
BiDir(A,B) = (d(A,B), d(B,A)).
\end{equation}
Hence, for any pair of networks A and B, we compute a two dimensional or bi-directional distance score in order to compare their modular structure. 

This method has three distinct advantages over previous community or partition comparison methods. First, it can be adapted to any community detection algorithm that is based on optimizing an objective function. Second, it goes beyond simply comparing the resulting partition arising from a community detection algorithm but accounts for the underlying network structure, specifically the strength of the connections within and between communities. This means, for example, that our method can capture differences between networks with similar partitions but where one of them might have a more defined or robust community structure. Third, it can also identify cases in which very dissimilar partitions hide the fact that the community structure of both networks is relatively similar. A network might be well suited to multiple distinct partitions - for example, when there are nested communities - all of which will score highly in the objective function. Unlike other methods, our framework will pick up this signal. 

\begin{figure}[!t]
    \centering
    \includegraphics[width=1\linewidth]{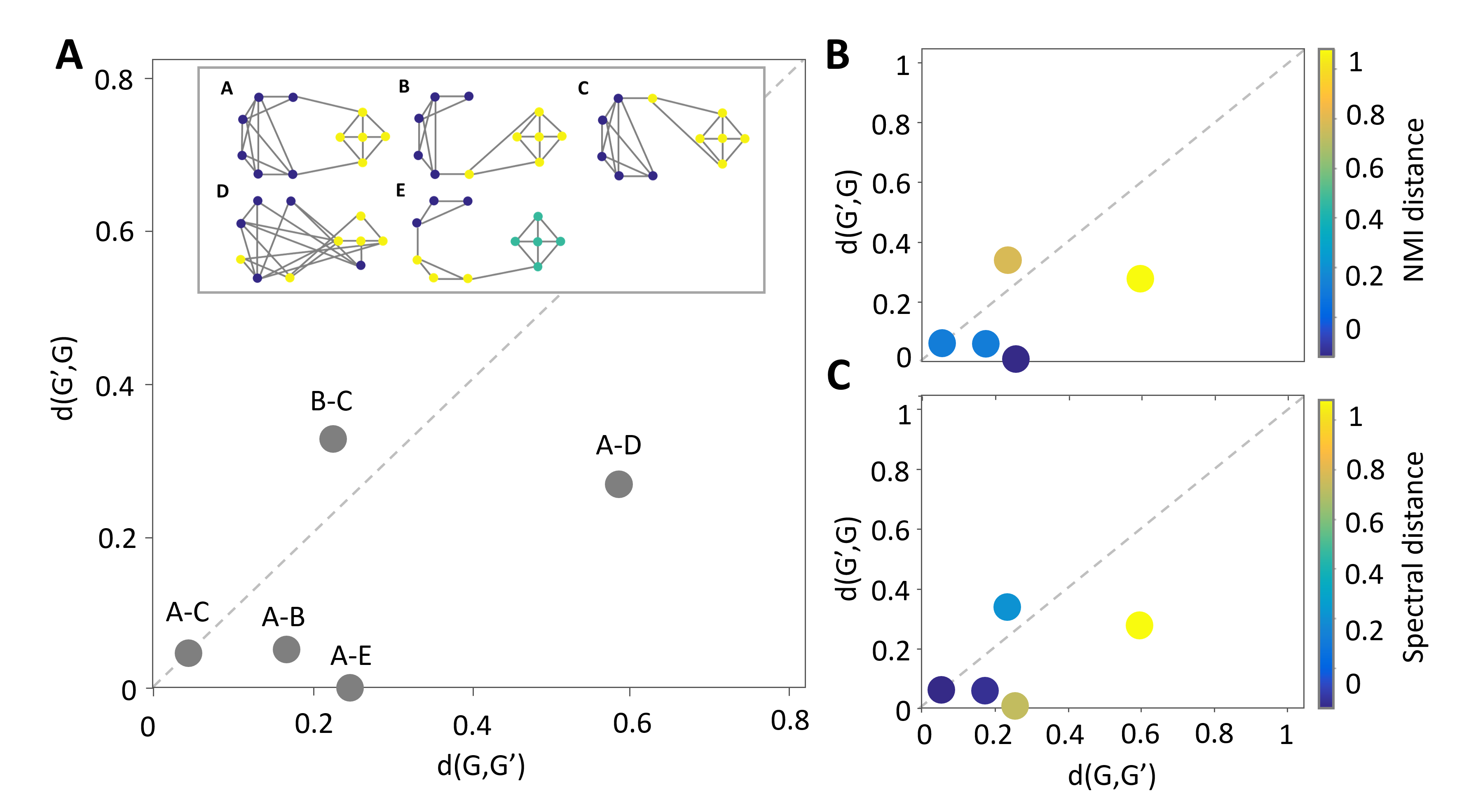}
    \caption{Using the same networks as Figure 1 (shown in the inset of subfigure \textbf{A}), we compare the same pairs using the BiDir distance, shown in \textbf{A}. We notice that distance between the pair A-C is small, as demonstrated by a BiDir value very close to the origin and identity line. Similarly, if we project the communities of A onto network E, we obtain a very good fit and a near zero distance (as shown by the point on the x axis). This is a consequence of the fact that the communities of E are a refinement of the ones in A. This does not hold true in the other direction, i.e., if we project the communities of E onto A, as shown by a non-zero distance of the point to the y axis. On the right, in \textbf{B} and \textbf{C}, we can see how rankings derived from other metrics, such as NMI and spectral distance, do not fully capture these features. For NMI, the A-E pair is similar, but not so for the spectral distance.}
    \label{ToyMethod}
\end{figure}

In Figure \ref{ToyMethod} we illustrate the key features of the BiDir distance using the toy networks from Figure \ref{ToyMotivation}. We use modularity optimization as the community detection algorithm. As outlined above, we obtain not one but two values for each pair of networks, which are plotted against each other in two-dimensional space in sub-figure A. 

The first thing to notice is that the two components of our distance convey distinct pieces of information. Points that are positioned away from identity line (diagonal) correspond to asymmetric behaviour, whereby one of the swapped partitions scores higher than the other. We can interpret this a case where the optimal partition of the first network is well-described by the modular structure of the second. The optimal partition of the second network, however, is not as compatible with the first. For example, we can see that the point corresponding to pair A-E is very close to the x axis. This indicates that if we project the communities of A onto network E, we obtain a very good fit. In fact, since the distance is near zero, the partition found on E is virtually as good a description of A's community structure as the communities found on A. Looking at the other direction, whereby we project the communities of E onto network A, do we not find as good a fit. By eyeballing the relevant networks, shown in the inset, we notice that the communities of E are a refinement of the communities of A, and thus we can think of them as being nested in the communities of A. This explains why the distance $d(E,A)$ is so small.

The second key observation is that our approach implicitly takes into account the network's connectivity patterns, and not just the partition, by making use of the objective or quality function. Notice that, unlike NMI (or like any other purely partition-based methodology), we differentiate between pairs A-B and A-C. 

They both are equally distant from the x axis, and so the optimal partition of A, when projected onto B or C, results in more or less the same modularity score. The community structure of C, however, scores better than B when projected onto network A. Again, looking at two networks we can understand why this is the case. 

In the former case, when we project the partition of A onto network B and C, the partition misplaces a single node (relative to the optimal partition of B and C). In the latter case, when we project the partition of B and C onto network A, we see that while both the partitions derived from B and C differ from A by a single node in the first case a node that is central to the blue community in A changes groups, while in the second it is the most peripheral node that does. Hence, the distance between A and C is smaller. 

This simple example shows how BiDir, by exploiting the objective function of the community detection algorithm, captures information on connectivity structure underlying the community structure of both networks, and is therefore capable of distinguishing cases which NMI and related methodologies would not be able to discriminate.

\subsection*{Synthetic networks}

Having introduced the BiDir distance, we will now construct several families of networks drawn from a stochastic block model (SBM) to further illustrate its features and test its performance. Specifically, we probe the effect of the strength of the community structure, and nestedness and overlap of communities, on the BiDir distance. 

Stochastic block models provide a good generating model for networks with a clearly defined community structure. First introduced to study the structure of social networks \cite{holland1983stochastic}, they have been widely adopted because they provide a straightforward way of generating modular networks as well as a simple benchmark for testing the statistical significance of communities. In a classical stochastic block model, $n$ nodes are partitioned into a set of $k$ communities $C=\{C_1,...,C_k\}$. A matrix $P$ of $k\times k$ encodes inter-community connection probabilities (the probability that nodes $v\in C_i$  and $u\in C_j$ share an edge is given by $P_{ij}$). Hence, we can directly specify the number of communities as well as the likelihood of the connection within and across communities.

It is possible to use SBMs to generate networks that exhibit a nested community structure. In this case, smaller communities are nested inside large communities. This is type of a structure that we observe in many real world networks. For example consider scientific collaboration networks: at a small scale we might identify individuals that belong or belonged to the same institution or research group, but these groups might also work together within a country or region, creating two levels of modular structure. 

\begin{figure}[!t]
    \centering
    \includegraphics[width=1\linewidth]{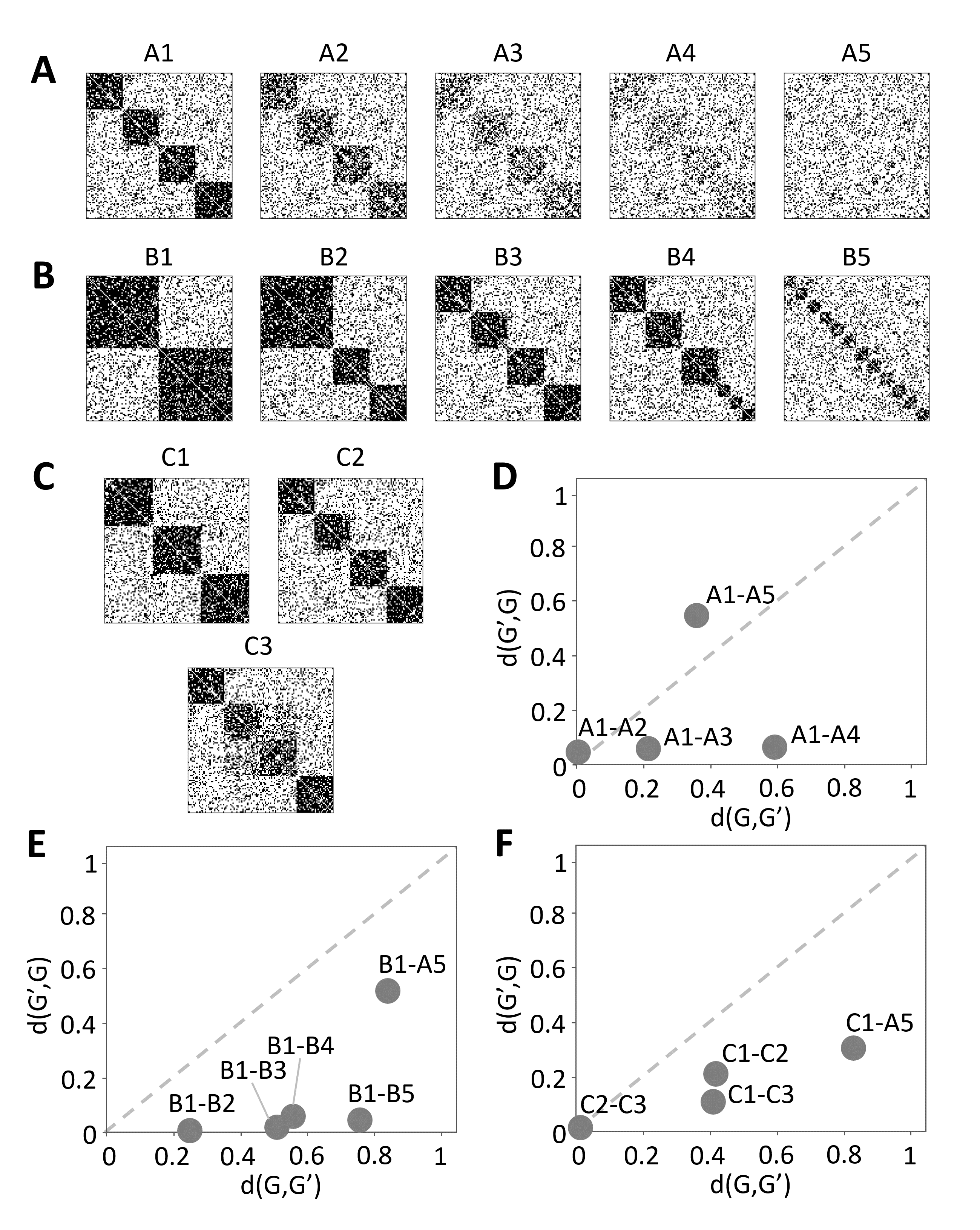}
    \caption{We illustrate our framework using three different families of SBMs. In family A all five networks contain the same communities, but the strength of in-group ties is sequentially decreasing. In subfigure D we see that the distances are zero whenever the original communities are detected, and increase with the weakening of the community structure. In family B each successive SBM is a refinement of the previous one, with blocks splitting along the way. So while B1 has two large communities, B3 has 4 communities, two for each of B1's communities. Notice that in subfigure E, the community structure of B1 describes well the structure of B2-B5, but that doesn't hold in the other direction. In the last family of SBMs C1 has three communities while C2 and C3 have four identical communities. But for C2, two of these communities can also be merged into one larger community. In subfigure F we can see that, while C2 and C3 have an identical optimal community structure, our method picks up that C1 is closer to C3 than to C2 because C3 can also be partitioned into 3 communities.}
    \label{SBM}
\end{figure}

We generate three families of SBM, as illustrated in Figure \ref{SBM}. Full details on the parameters used to generate these ensembles is provided in the SI. As before, modularity is used as the community detection algorithm. The distances shown correspond average BiDir distances over $1000$ instances of each SBM. 

The first family has four SBMs, all with the same community structure but with successively decreasing probability of in-group ties. As long as the communities can be recovered, their structure remains intact and thus the distance between them will be $0$. Once the communities can no longer be properly identified by our community detection algorithm, the distances increase. We see in subfigure D that when we project the community structure of A1 onto A2, A3 and A4, we obtain values close to zero (i.e., the points lie along the x axis). As expected, A1's community structure is very close to being optimal for A2, A3 and A4 as well. However, if we instead project the community structure detected for A2, A3 and A4 onto A1, the distance progressively increases as the community detection algorithm struggles to find the underlying or generating structure. The final SBM A5, where inter and intra community connection probabilities are equal, does not have a community structure and is thus most distant from A1.

The second family is composed of five SBMs, each a refinement of the previous one. In other words, B1 has two communities, B2 has three communities, obtained by splitting one of the communities of B1 into two, B3 has four communities and so on. We compare the first SBM B1 to the other four. We observe, as we would expect, that as we split the communities further, we increase the distance from the original network. More specifically, if we project the community structure of B1 onto B2-B5, we again obtain values close to zero (i.e., the points lie along the x axis). Hence the original partition B1 remains informative even after a few splits since the new communities are nested inside the original one. Looking in the other direction, however, the more dis-aggregate partitions describe progressively less well the structure of B1.

Our last family of SBMs is slightly more complicated. C1 has three communities, but none of them maps onto the communities in C2 or C3. C2 and C3 are both split into four identical communities, but for C3 two of these communities are also strongly connected between them, and thus C3 can also be reasonably partitioned into three communities. In subfigure F we can see that the $D(C2,C3) = (0,0)$ - since the partition of C3 into four communities has higher modularity than the partition into three, C2 and C3 the same optimal community structure. But when comparing them to C1, we see a different picture. C3 is closer to C1 because the middle community of C1 is close to the combined middle block of C3. C2 is further from C1 as the middle community of C1 overlaps two very distinct blocks in C2. We see that BiDir distance rightly identifies that although C2 and C3 have the same optimal partitions, their underlying community structure is markedly distinct.  

Hence, even when the optimal partitions of two networks appear to exhibit little overlap, our method will generate a low distance score when there exists an alternative and almost optimal partition that is similar. It is well known \cite{fortunato2010community} that an exhaustive optimization of modularity is impossible due to the high number of possible partitions, and furthermore, very different partitions might obtain similar near optimal scores. Therefore looking exclusively at the resulting partitions is likely to miss important structural similarities that are nevertheless captured by the BiDir distance whenever these two distinct partitions produce similar scores for the optimization function.

\subsection*{Application to inter-industry labour flow networks}

In this section,  we compare the modular structure of five European Skill-Relatedness Networks (SRN) using the BiDir distance. Recall that, although it has previously been argued that the SRN is universal, we expect to find variation across the modular structure of different countries' SRN. Understanding this variation will unveil how structural characteristics of labour markets impact on industrial diversification paths. 

An SRN is constructed as follows. Two industries are considered skill-related if there is a larger number of job switches between them compared to what would be expected at random. Formally, if $\phi_{ij}$ is the number of job switches between industry $i$ and $j$ within a given time period, then the skill-relatedness is given by
\begin{equation}
   S_{ij}=\frac{\phi_{ij}/\sum_j \phi_{ij}}{\sum_i\phi_{ij}/\sum_{ij}\phi_{ij}}.
\end{equation}
The matrix is then made symmetric by averaging with its transpose and re-scaling the values so that they range from $-1$ to $1$ to give the adjacency matrix:
\begin{equation}
    A_{ij}=\frac{S_{ij}+S_{ji}+2}{S_{ij}+S_{ji}-2}.
\end{equation}
Note that values that are greater than $0$ indicate that the number of job switches is greater than what would be expected at random under a null model (specifically, the configuration model \cite{molloy1995critical}). We therefore conserve only positive values of this matrix. The reader is referred to Neffke \emph{et al} \citep{Neffke2013SkillRelatedness} for a detailed discussion of these methodological considerations. 

Here we compare the SRNs for Germany \citep{neffke2011}, Ireland \citep{o2019modular}, the Netherlands \citep{diodata2014}, the United Kingdom and Sweden \citep{neffke2013skill}. In all cases, the industry classification corresponds to the 4-digit NACE 1.1 industry classification\footnote{Due to data availability, the United Kingdom SRN was constructed using the 4-digit NACE 2 classification and then converted to the NACE 1.1 classification. Readers are referred to the SI for a more detailed discussion regarding this conversion.}. The graph intercept of these networks (with the exception of the UK, see footnote below) is used to ensure that all networks have the same size and are node-aligned - a requirement of the BiDir metric.  

\begin{figure}[!t]
    \centering
    \includegraphics[width=1\linewidth]{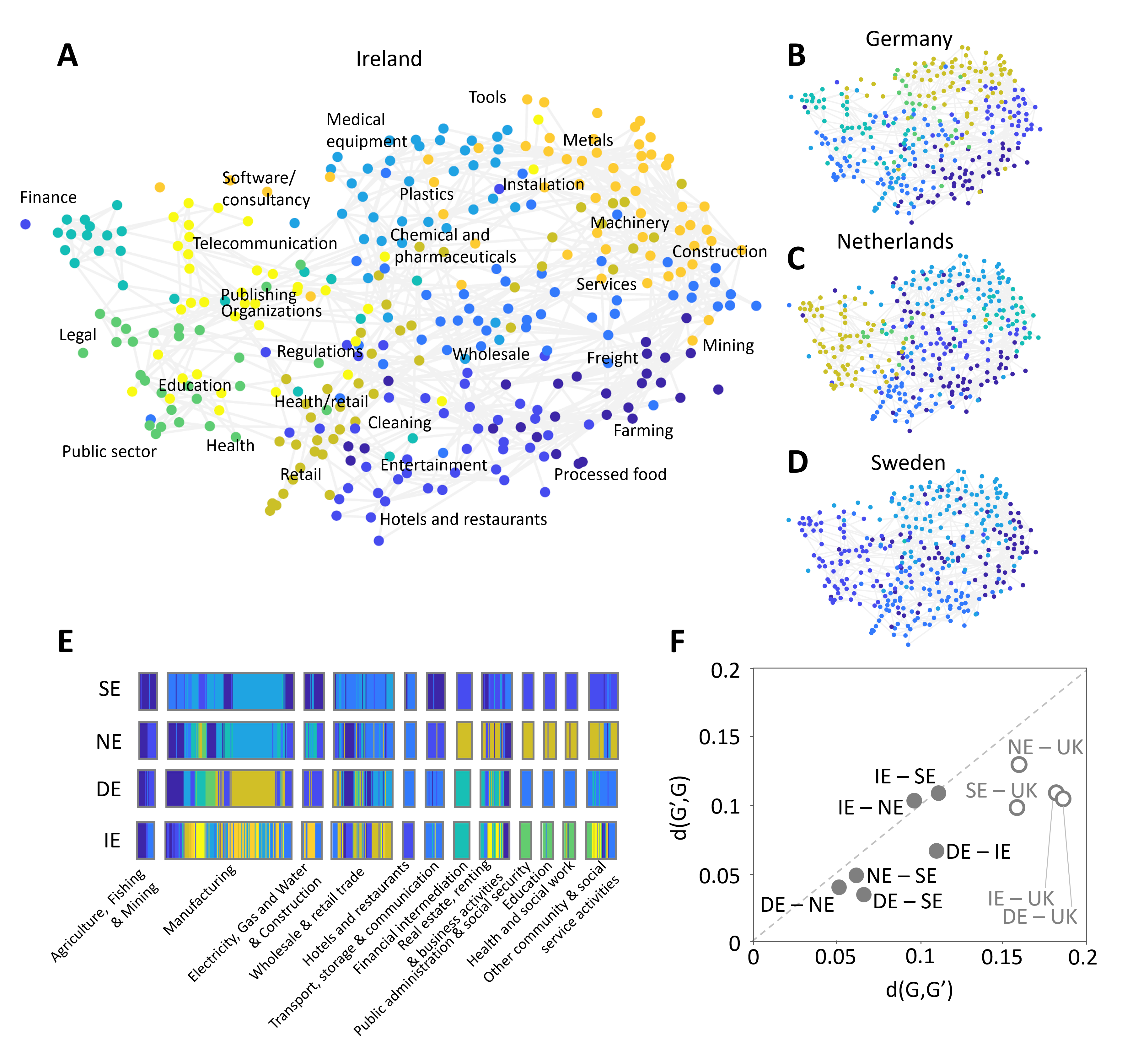}
    \caption{The pairwise comparison of the modular structure of five countries' SRNs, namely: Ireland (IE), Germany (DE), the Netherlands (NE), the United Kingdom (UK) and Sweden (SE). In (A) we show the Irish SRN from O'Clery \emph{et al} \cite{o2019modular}. The nodes represent industries and edges the skill-relatedness between the two corresponding industries. Nodes are coloured according to their community (found using the modularity optimization algorithm). The node layout is based on a spring algorithm called `Force Atlas' in Gephi. Labelling is added to indicate the general position of sectors within the network. Using the same layout we also show communities detected for Germany, the Netherlands and Sweden in sub-figure (B) to (D). In (E) we highlight the overlap of the community structure of each country with the official industrial sector classification. Industries are ordered and grouped by their 1-digit NACE 1.1 sector along the x-axis. For each country, industries are coloured according to community membership. In (F) we show the resulting BiDir distance from the pairwise comparison of the different SRNs. Notice that the German, Swedish and Dutch SRNs have relatively similar modular structure, while the Irish SRN's modular structure is less similar.}
    \label{SRNetCom}
\end{figure}

In Figure~\ref{SRNetCom} (A) we show a visualisaton of the Irish SRN from O'Clery \emph{et al} \citep{o2019modular}. Each node in the network represents an industry and each edge the skill-relatedness between its two corresponding industries. The node layout is based on a spring algorithm called `Force Atlas' in Gephi, where more skill-related industries are positioned closer together. We have added labelling to indicate the general position of sectors in the network. We observe that service-orientated industries and government activities tend to be located on the left hand side of the network, whereas heavy goods, construction, manufacturing and agricultural sectors dominate the right. Retail (bottom) and business (top) activities lie in between.

The modular structure of the network is shown via the node colouring. The communities were detected using modularity \cite{newman2006modularity}. Specifically, the partition maximizing the modularity function from $10 000$ iterations of the Louvain algorithm is adopted. Using the same layout, we visualise the communities found for Germany, the Netherlands and Sweden on the right. 

Figure \ref{SRNetCom} (E) highlights the overlap of the community structure of each country with the official industrial sector classification. Industries are ordered and grouped by their 1-digit NACE 1.1 sector along the x-axis. For each country, industries are coloured according to community membership, corresponding to the node colouring in sub-figure (A) to (D). Hence, blocks of colour within a sector indicates a close correspondence between a community and a sector. Conversely, a diversity of colours within a sector indicates a community structure that is quite distinct from the official classification.   

This figure provides insight into how the modular structure varies across the different SRNs. First, we observe that the Irish SRN has more communities than the other 3 networks. The Irish SRN has 9 communities, the Netherlands and Germany have 7 communities, while Sweden only has 5 communities. We can also see that industries within certain sectors are always clustered together across all four networks (as indicated by solid colour blocks), such as financial intermediation, public administration and social security, agriculture and the hotel and restaurant sectors. These communities show universal structure in inter-industry skill-sharing. Furthermore, insights such as the increased subdivision of the manufacturing sector in Ireland compared to other SRNs can also be observed. 

We now quantify the distance between these networks using our BiDir distance. The pairwise comparison of the various SRNs is shown in Figure \ref{SRNetCom} (F)\footnote{The United Kingdom's SRN is has fewer industries compared to the other four countries, and also much more sparse. This is because it was constructed from a longitudinal survey including just 1 per cent of UK workers. Hence, when we compare the UK to the other countries, we compare the subgraphs induced by the set of overlapping nodes (and corresponding edges) in each case. A more detailed comparison of the UK to the other countries is included in the SI.}. We observe that Germany and the Netherlands are the closest in terms of modular structure with the smallest distance, and both are similar to Sweden. Ireland appears to have a very different modular structure compared to all other countries. Comparison of the Irish SRN and the German SRN uncover an asymmetric relationship. If we project the German communities onto the Irish SRN, we obtain a good fit - but not vice versa. Visual inspection of subfigures (A), (B) and (E) confirms that the communities of Ireland are somewhat nested inside the communities of Germany. For example, the manufacturing and social service sectors consists of many small communities in Ireland compared to just a few larger communities in Germany. Hence, inter-industry flows, and thus knowledge diffusion and skill-sharing that constrains development paths, are more fragmented in Ireland. From an Irish policy perspective, the German SRN might be informative in the design of smart interventions. For example, policies could be developed to facilitate and encourage worker mobility between sectors in, for example, manufacturing that are highly interconnected in Germany, but not yet in Ireland. 

Hence, our BiDir distance reveals a fairly universal modular structure in terms of inter-industry flows and skill-sharing across a set of Northern European countries. The directionality of our distance metric enables us to pick up nuanced differences between SRNs, and uncover unseen potential linkages which are key to policy efforts to generate regional industrial growth and diversification potential. 

\subsection*{Different optimization functions}

The flexibility of BiDir, in terms of its adaptability to a class of community detection algorithms (with the use of an objective function), is advantageous in that it facilitates the comparison of networks across a wide variety of applications. This is because different optimization functions define communities differently and will therefore often detect different partitions even on a single network. Thus far, we have illustrated the use of BiDir when using the modularity optimization function \citep{newman2008physics} as our objective function. Here, we illustrate BiDir distances obtained when comparing the modular structure of two networks using three different optimization functions. Specifically, we consider InfoMap \citep{Rosvall2009}, Markov stability \citep{delvenne2010stability} and modularity functions. Specifically, we illustrate the degree to which properties of an optimization function are inherited by and reflected in our distance measure.

\begin{figure}[!t]
    \centering
    \includegraphics[width=0.95\linewidth]{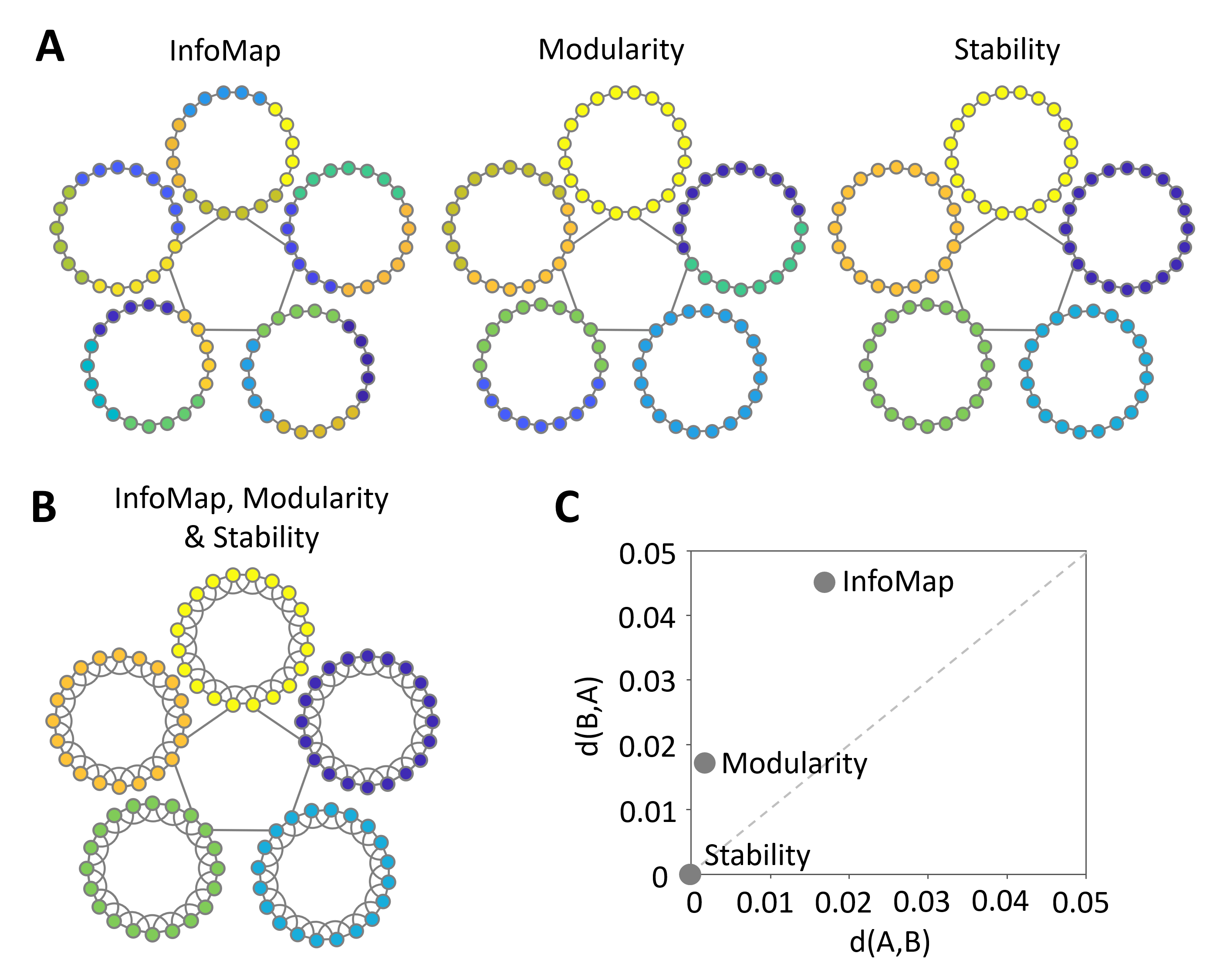}
    \caption{We illustrate how the BiDir distance varies when using three different optimization functions, namely: Modularity, Infomap and stability. In (A) and (B) we illustrate network A and B, respectively, with node colouring corresponding to the communities detected. In (C) we show the BiDir distances obtained when comparing the modular structure of network A and B using each of the different optimization functions. A larger distance is obtained when using the modularity and Infomap function compared to the stability function. This is because the first two functions suffer from a 'field-of-view' limit and therefore over-partition network A. In general, the BiDir distance inherits the features of the optimization function deployed.}
    \label{Flexibility}
\end{figure}

We compare the modular structure of networks A and B illustrated in Figure~\ref{Flexibility} (A) and (B), respectively. Network A is a ring-of-rings and is an example of a network with non-clique-like communities (communities that have a high average diameter). The network is taken from Schaub \emph{et al} \citep{Schaub2012}. Network B is a ring-of-circulant graphs $C_n \langle 1,2 \rangle $ \footnote{A circulant graph is a graph of order $n$ in which the $i^{\text{th}}$ vertex is adjacent to the $(i+j)^{\text{th}}$ and $(i-j)^{\text{th}}$ graph vertices for each $j$ in a so-called connection set. A circulant graph of order $n$ and with connection set $\{1,2\}$ is denoted as $C_n\langle 1,2 \rangle$.}. It has a similar community structure as network A, however the clique-like structure of each community is enhanced (the average diameter of each community is decreased) by the addition of intra-community edges.

For modularity and InfoMap the partition is obtained by applying the Louvain- and internal InfoMap heuristic $1000$ times respectively, and choosing the partition which either maximises the modularity or minimises the InfoMap function. Similarly, in the case of the multi-resolution stability algorithm (where time is the intrinsic resolution parameter), for each time a resulting partition is obtained by applying the Louvain heuristic $1000$ times. For each time resolution, we identity the partition that maximizes the stability function. In order to choose the partition corresponding to the optimal resolution, we calculate the mean variation of information\footnote{The variation in information is a metric used to measure the robustness of our partitions. The variance of information calculates the amount of information (in an information-theoretical sense) two partitions share. Two partitions that are similar will exhibit a low variation of information. For each resolution a pairwise variance of information is calculated for each partition obtained from the Louvain algorithm. The average of these is then the mean variation of information at the resolution. If this value is low, it shows that the detected partitions are similar, and therefore more robust \citep{meilua2007comparing}.} (VI) \citep{meilua2007comparing} at each resolution, and choose the partition corresponding to the resolution with lowest VI. We consider the full range of partitions obtained at different resolutions in the next section.

The resulting partitions are visualised via node colouring. As noted by Schaub \emph{et al} \citep{Schaub2012}, we observe that both InfoMap and modularity over-partition network A (each ring is partitioned into multiple communities). This can be explained by the `field-of-view' limit of which both of these functions suffer \citep{Good2010}. The `field-of-view limit' is an upper limit on the effective diameter of the communities that can be detected with one-step dynamical community detection techniques \citep{Schaub2012}. As the rings in network A have high diameters (non-clique like structure) both these algorithms over partition the rings, creating communities with smaller diameters. However, this over-partitioning is not observed in the case of network B. This is because each of the circulants in network B have a smaller diameters and the optimization functions are able to partition them into their own community. As the Markov stability function is a multi-step dynamical community detection algorithm it does not suffer from the field-of-view limit \citep{Schaub2012}, and obtains a community structure consistent with what we would expect for both network A and B. 

In Figure~\ref{Flexibility} (C), we illustrate the BiDir distances (comparing network A and B) obtained when adopting each of these optimization functions. As expected, Infomap and modularity result in the largest distances. Both of these distances are also highly asymmetric. We find that if we project the communities of B onto network A, we obtain a good fit - but not vice versa. The BiDir distance therefore recognises the subdivided communities found using modularity and Infomap. As InfoMap results in the most severe over-partitioning of the network, it results in the largest and most asymmetric distance. 
As the Markov stability algorithm does not suffer from the `field-of-view limit' it obtains the same community structure for network A and B, and results in a distance of $(0,0)$. The different distances obtained using the different optimization functions illustrates that the BiDir distance inherits the properties of the community detection algorithm used. 

\subsection*{Multi-scale optimization functions}

Although most well-known community detection algorithms seek to obtain a single node partition, in many cases it is more natural to analyse a range of partitions - from many small communities to a few large communities - when investigating the modular structure of a network. Such hierarchical structure can be informative, revealing layers of organisation. For example, consider a university friendship network. Larger communities may reveal institutional structures such as departments or colleges, while smaller communities might capture friendship or social circles. 

Hence, when comparing the modular structure of two networks, one may want to consider not a single partition but a range of partitions for each network. This approach can reveal, for example, distinct distances corresponding to particular scales, and consequently uncover the resolution at which the two networks are most similar. For example, two networks may be quite similar in terms of their modular structure at a coarse scale, but distinct at a finer or more dis-aggregate scale. Taking further advantage of the flexibility of our method, here we illustrate the use of the BiDir distance for a multi-resolution optimization function. 

To illustrate how we adapt the BiDir distance to this case, again consider two networks, A and B. We compute the ratio of A's $F$-score under B's optimal partition found at resolution $\beta$ to its $F$-score under its own optimal partition found at resolution $\alpha$, and vice versa. Formally, 
\begin{equation}\label{comss2}
d(A_\alpha,B_\beta) = 1-\frac{F(A,P_{B_\beta})}{F(A,P_{A_\alpha})},
\end{equation}
where $P_{A_\alpha}$ is the optimal partition of network A obtained at resolution $\alpha$ and $P_{B_\beta}$ is the optimal partition of network B obtained at resolution $\beta$. Hence, the final BiDir distance corresponds to: 
\begin{equation}\label{comss2b}
D(A_\alpha,B_\beta) = (d(A_\alpha,B_\beta),d(B_\beta,A_\alpha)).
\end{equation}
In practice, we calculate this distance for a range of $\alpha$ and $\beta$ values such that small values of $\alpha$ or $\beta$ correspond to fine partitions with many small communities, while larger values correspond to fewer larger communities\footnote{The range of $\alpha$ and $\beta$ is network specific and chosen by the user.}. 

\begin{figure}[!t]
    \centering
    \includegraphics[width=0.95\linewidth]{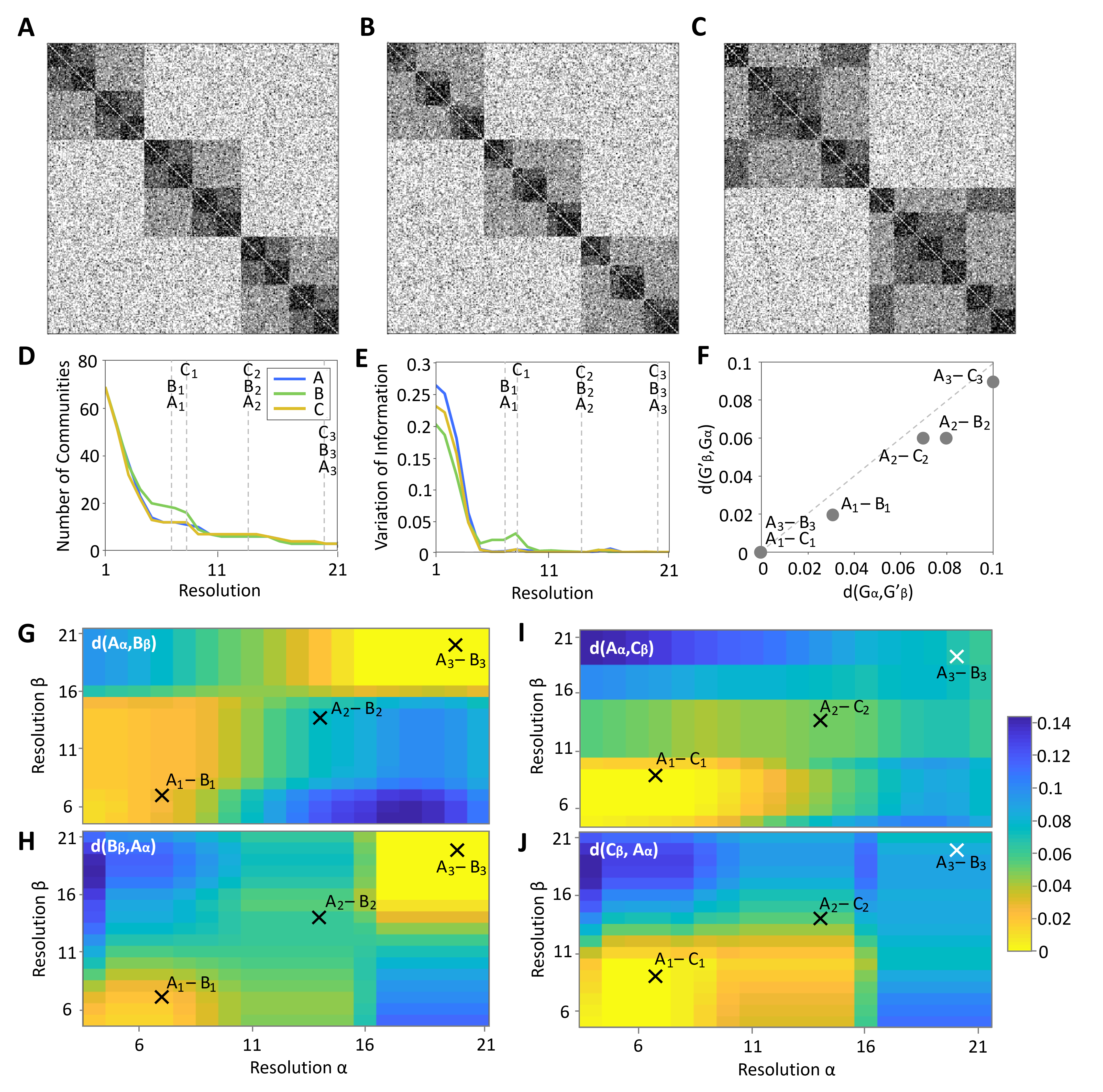}
    \caption{We illustrate the use of the BiDir distance to compare the modular structure of three nested SBMs. (A--C): We visualize a single network realisation for each SBM. Network A and B share a similar coarse-level community structure with 3 equally sized communities, while A and C share a similar granular-level community structure with 12 equally sized communities. In (D and E): We show the number of communities and the mean variation of information obtained using the Markov stability algorithm for each of the SBMs for a range of resolutions. For each network, we highlight three robust partitions corresponding to distinct scales. (F): We show the BiDir distances obtained when comparing these partitions. (G and H): We illustrate the BiDir distances obtained when comparing networks A and B and networks A and C across the full range of resolutions. In general, we observe that network A and B's modular structure is most similar (e.g., the distance is smallest) at a coarse scale with few large communities, while networks A and C are most similar at a dis-aggregate or finer scale.} 
    \label{Res}
\end{figure}

To illustrate this approach, we compare three families of networks generated by hierarchical SBMs that share modular structure at three different scales. All three SBMs consist of 300 nodes (see SI for more details). In Figure ~\ref{Res} A--C, we visualise the adjacency matrix for a single network generated by each SBM. Note that A and B share a coarse-level modular structure of three equally size communities. However, at a more granular-level, their community partitions are more dissimilar. On the other hand, network A and C share a granular-level modular structure of 12 equally sized communities.
 
Next, we use the multi-resolution Markov stability community detection algorithm to detect a range of partitions for each of the networks. Recall, that this algorithm is based on a simple random walk model \cite{delvenne2010stability, lambiotte2008dynamics, lambiotte2011flow}. The key idea behind this method is that it sets a walker to roam on a network - jumping from node to node with probability proportional to the edge weight. If the walker gets trapped in a region of the network (a group of nodes) for a prolonged period this corresponds to a group of densely connected nodes which form a community. Here, the Markov 'time' is used as the resolution parameter. Intuitively, if we let a walker roam for longer periods on the network, the walker will detect larger and larger communities. Therefore, by varying the Markov times we detect communities on a range of scales: from many small communities to few large communities. More detail on this algorithm can be found in Lambiotte \emph{et al} \cite{lambiotte2011flow}. 

We adopt the Louvain algorithm \cite{Blondel2005} as our search algorithm to find the optimal node partition with respect to our objective function (the stability function \cite{delvenne2010stability}) at each resolution. We use variance of information (across a large number of realisations of the Louvain algorithm) to assess which of these partitions is the most robust. In sub-figure D and E, we display the number of communities and the mean variation of information for each resolution\footnote{Note that for all results, we calculate the average values obtained from generating $100,000$ network instances of each SBM and running the Markov stability algorithm with $1,000$ iterations of the Louvain heuristic on each network instance generated.}. We highlight three resolutions that are both robust, with low variation of information, and correspond to three distinct scales of community structure.  

First, we use the BiDir distance to compare these chosen partitions in the standard manner, as shown in (F). We clearly observe that the distance varies depending on the resolutions chosen. For example, partitions A$_3$ and B$_3$, corresponding to larger scale coarse partitions of both A and B, are much more similar than A$_1$ and B$_1$ (fine) or A$_2$ and B$_1$ (moderate). 

Next, we compare the modular structures of network A and B across the full range of resolutions. Figure~\ref{Res} (G and H) shows values obtained for different $(\alpha, \beta)$ resolution pairs. As our distance is two-dimensional, (G) corresponds to direction $d(A(\alpha),B(\beta))$ and (H) corresponds to direction $d(B(\beta),A(\alpha))$. Consistent with our example above, we observe that A and B's modular structure is most similar (e.g., the distance is smallest) at a coarse scale with few large communities. As we increase the granularity of the community structure, the modular structure of the networks becomes more distinct (at a moderate scale) and then again more similar (at a fine scale). 

We also observe an asymmetric relationship with respect to modular distance. For small $\alpha$ and moderate $\beta$ we observe that $d(A(\alpha),B(\beta)) < d(B(\beta),A(\alpha))$. Hence, if we project the communities of network B onto network A, we obtain a good fit - but not vice versa. In this region network B has fewer and larger communities (9 communities) than network A (which has 12 communities). Thus, our metric picks up that the stability of both partitions are similar with regard to network A. 

Conversely, when we compare networks A and C in (I and J), we observe that the smallest distance is obtained at a fine resolution corresponding to detection of smaller communities. As the resolution increases, and larger communities are detected, the modular structure of the networks becomes less similar. Here we observe a stronger symmetric relationship between the two dimensions of the BiDir distance. 

\section*{Discussion}

Comparing community structure across networks remains a challenging task. The majority of current methodologies equate the community structure to the graph partition, but this approach ignores key topological information. While we also compare two partitions, by assessing the fit of one partition to the topology of the other network (and vice versa), the bi-directional distance we propose incorporates information on the network structure. This approach enables us to pick up similarities and differences in the underlying network topologies not captured in the partitions alone. Agnostic about the community detection quality function used within a broad class, this approach can be deployed across a wide range of applications. 

The BiDir distance assumes that the partition assigned to each network is optimal (under some chosen community detection quality function). While this might appear like a strong assumption, the BiDir distance is well suited to handle cases where there exist alternative partitions that are close to optimal, since they all are assigned similar scores by the quality function. This feature enables our method to detect topological similarities that are not captured by (potentially dissimilar) partitions.

The multi-dimensionality of BiDir means that it is neither a metric nor a quasi-metric in a formal sense. It is therefore important to be careful using it for certain applications. For example, when bench-marking different community detection algorithms using ground truth data, it will not produce a unique ordering. Furthermore, the scores given by BiDir are only comparable within a set of related networks (networks that are of the same size and exactly node-aligned). Similarly, since BiDir inherits the properties of the community detection algorithm used, the partitions to be compared should be derived from the same community detection algorithm. Further work is required to investigate if it is possible to normalize BiDir to enable more general comparisons of scores for different classes of networks.

A key aspect of community detection is that the specific algorithm is chosen according to whether structural, dynamic or other features are of interest. Since we use the quality function in our method, we retain this focus on the features of interest. However, our method is limited to algorithms that optimise an objective function. Extensions to other algorithms could be explored in future work. For example, if one was interested in comparing two networks using an SBM model, one could compare the statistical significance of two alternative SBM models (e.g., using a posterior odds ratio) instead of using an $f$-score. 

Finally, our measure identifies the presence of similar modular structure between network pairs. However, when it comes to applications, often more information is required to understand 'where' in these networks similarities (or differences) occur. This was clearly illustrated in the comparison of the inter-industry labour flow networks, where a finer investigation of community overlap was used to shed light on sectors with either conserved or distinct patterns of labour flows across countries. The adaption of our bi-directional distance to a finer community-level distance is a potentially fruitful future research endeavour.

\bibliographystyle{plain}
\bibliography{references.bib} 

\begin{thebibliography}{10}

\bibitem{barabasi1999mean}
Albert-L{\'a}szl{\'o} Barab{\'a}si, R{\'e}ka Albert, and Hawoong Jeong.
\newblock Mean-field theory for scale-free random networks.
\newblock {\em Physica A: Statistical Mechanics and its Applications},
  272(1-2):173--187, 1999.

\bibitem{Barahona2003}
M.~Barahona and L.~M. Pecora.
\newblock Synchronization in small-world systems.
\newblock {\em Physical Review Letters}, 89.5:54101, 2002.

\bibitem{barahona2002synchronization}
Mauricio Barahona and Louis~M Pecora.
\newblock Synchronization in small-world systems.
\newblock {\em Physical review letters}, 89(5):054101, 2002.

\bibitem{Blondel2005}
V.D. Blondel, J.M. Hendrickx, A.~Olshevsky, and J.N. Tsitsiklis.
\newblock Convergence in multiagent coordination, consensus, and flocking.
\newblock {\em 44th IEEE Conference on Decision and Control}, pages 2996--3000,
  2005.

\bibitem{chakraborty2017metrics}
Tanmoy Chakraborty, Ayushi Dalmia, Animesh Mukherjee, and Niloy Ganguly.
\newblock Metrics for community analysis: A survey.
\newblock {\em ACM Computing Surveys (CSUR)}, 50(4):1--37, 2017.

\bibitem{cranmer2011inferential}
Skyler~J Cranmer and Bruce~A Desmarais.
\newblock Inferential network analysis with exponential random graph models.
\newblock {\em Political analysis}, 19(1):66--86, 2011.

\bibitem{csafordi2016effect}
Zsolt Cs{\'a}fordi, L~L{\'a}szl{\'o}, Bal{\'a}zs Lengyel, and
  K{\'a}roly~Mikl{\'o}s Kiss.
\newblock The effect of labor flows, ownership and skill-relatedness on firm
  productivity.
\newblock In {\em Proceedings of International Academic Conferences}, number
  4006263. International Institute of Social and Economic Sciences, 2016.

\bibitem{Danon2005}
L.~Danon, A.~Diaz-Guilera, J.~Duch, and A.~Arenas.
\newblock {Comparing community structure identification}.
\newblock {\em {Journal of Statistical Mechanics: Theory and Experiment}},
  2005(09):P09008, 2005.

\bibitem{delvenne2010stability}
J-C Delvenne, Sophia~N Yaliraki, and Mauricio Barahona.
\newblock Stability of graph communities across time scales.
\newblock {\em Proceedings of the National Academy of Sciences},
  107(29):12755--12760, 2010.

\bibitem{diodato2018network}
D~Diodato.
\newblock A network-based method to harmonize data classifications.
\newblock Technical report, Utrecht University, Department of Human Geography
  and Spatial Planning, 2018.

\bibitem{diodata2014}
D~Diodato and A~B Weterings.
\newblock {The resilience of regional labour markets to economic shocks:
  Exploring the role of interactions among firms and workers}.
\newblock {\em {Journal of Economic Geography}}, 15(4):723--742, 2014.

\bibitem{diodato2014resilience}
Dario Diodato and Anet~BR Weterings.
\newblock The resilience of regional labour markets to economic shocks:
  Exploring the role of interactions among firms and workers.
\newblock {\em Journal of Economic Geography}, 15(4):723--742, 2014.

\bibitem{donnat2018tracking}
Claire Donnat and Susan Holmes.
\newblock Tracking network dynamics: A survey of distances and similarity
  metrics.
\newblock {\em arXiv preprint arXiv:1801.07351}, 2018.

\bibitem{dunne2002network}
Jennifer~A Dunne, Richard~J Williams, and Neo~D Martinez.
\newblock Network structure and biodiversity loss in food webs: robustness
  increases with connectance.
\newblock {\em Ecology letters}, 5(4):558--567, 2002.

\bibitem{farjoun1994beyond}
Moshe Farjoun.
\newblock Beyond industry boundaries: Human expertise, diversification and
  resource-related industry groups.
\newblock {\em Organization Science}, 5(2):185--199, 1994.

\bibitem{fortunato2010community}
Santo Fortunato.
\newblock Community detection in graphs.
\newblock {\em Physics Reports}, 486(3-5):75--174, 2010.

\bibitem{frenken2007}
R.~Frenken, F.~{Van Oort}, and T.~Verburg.
\newblock Related variety, unrelated variety and regional economic growth.
\newblock {\em Regional Studies}, 41(5):685--697, 2007.

\bibitem{girvan2002community}
Michelle Girvan and Mark~EJ Newman.
\newblock Community structure in social and biological networks.
\newblock {\em Proceedings of the national academy of sciences},
  99(12):7821--7826, 2002.

\bibitem{Good2010}
B.H. Good, Y.A. De~Montjoye, and A.~Clauset.
\newblock Performance of modularity maximization in practical contexts.
\newblock {\em {Physical Review E}}, 81(4):046106, 2010.

\bibitem{hagen1992new}
Lars Hagen and Andrew~B Kahng.
\newblock New spectral methods for ratio cut partitioning and clustering.
\newblock {\em IEEE transactions on computer-aided design of integrated
  circuits and systems}, 11(9):1074--1085, 1992.

\bibitem{hamming1950error}
Richard~W Hamming.
\newblock Error detecting and error correcting codes.
\newblock {\em The Bell system technical journal}, 29(2):147--160, 1950.

\bibitem{hausmann2007}
Ricardo {Hausmann}, Jason {Hwang}, and Dani {Rodrik}.
\newblock What you export matters.
\newblock {\em Journal of Economic Growth}, 12(1):1--25, 2007.

\bibitem{hidalgo2007product}
C{\'e}sar~A Hidalgo, Bailey Klinger, A-L Barab{\'a}si, and Ricardo Hausmann.
\newblock The product space conditions the development of nations.
\newblock {\em Science}, 317(5837):482--487, 2007.

\bibitem{holland1983stochastic}
Paul~W Holland, Kathryn~Blackmond Laskey, and Samuel Leinhardt.
\newblock Stochastic blockmodels: First steps.
\newblock {\em Social networks}, 5(2):109--137, 1983.

\bibitem{jaccard1912distribution}
Paul Jaccard.
\newblock The distribution of the flora in the alpine zone. 1.
\newblock {\em New phytologist}, 11(2):37--50, 1912.

\bibitem{jaffe1989characterizing}
Adam~B Jaffe.
\newblock Characterizing the ``technological position'' of firms, with
  application to quantifying technological opportunity and research spillovers.
\newblock {\em Research Policy}, 18(2):87--97, 1989.

\bibitem{jeong2000large}
Hawoong Jeong, B{\'a}lint Tombor, R{\'e}ka Albert, Zoltan~N Oltvai, and A-L
  Barab{\'a}si.
\newblock The large-scale organization of metabolic networks.
\newblock {\em Nature}, 407(6804):651--654, 2000.

\bibitem{krause1997soziale}
Ulrich Krause.
\newblock Soziale dynamiken mit vielen interakteuren. eine problemskizze.
\newblock {\em Modellierung und Simulation von Dynamiken mit vielen
  interagierenden Akteuren}, 3751:2, 1997.

\bibitem{lambiotte2008dynamics}
R~Lambiotte, JC~Delvenne, and M~Barahona.
\newblock Dynamics and modular structure in networks.
\newblock {\em arXiv preprint arXiv:0812.1770}, 2008.

\bibitem{lambiotte2011flow}
Renaud Lambiotte, Roberta Sinatra, J-C Delvenne, Tim~S Evans, Mauricio
  Barahona, and Vito Latora.
\newblock Flow graphs: Interweaving dynamics and structure.
\newblock {\em Physical Review E}, 84(1):017102, 2011.

\bibitem{masuda2017random}
Naoki Masuda, Mason~A Porter, and Renaud Lambiotte.
\newblock Random walks and diffusion on networks.
\newblock {\em Physics reports}, 716:1--58, 2017.

\bibitem{meilua2007comparing}
Marina Meil{\u{a}}.
\newblock Comparing clusterings—an information based distance.
\newblock {\em Journal of Multivariate Analysis}, 98(5):873--895, 2007.

\bibitem{milo2002network}
Ron Milo, Shai Shen-Orr, Shalev Itzkovitz, Nadav Kashtan, Dmitri Chklovskii,
  and Uri Alon.
\newblock Network motifs: simple building blocks of complex networks.
\newblock {\em Science}, 298(5594):824--827, 2002.

\bibitem{molloy1995critical}
Michael Molloy and Bruce Reed.
\newblock A critical point for random graphs with a given degree sequence.
\newblock {\em Random Structures \& Algorithms}, 6(2-3):161--180, 1995.

\bibitem{mucha2009communities}
Peter~J Mucha, J~Onnela, and MA~Porter.
\newblock Communities in networks.
\newblock {\em Notices of the American Mathematical Society}, 56:1082--1097,
  2009.

\bibitem{Neffke2013SkillRelatedness}
F~Neffke and M~Henning.
\newblock Skill relatedness and firm diversification.
\newblock {\em Strategic Management Journal}, 34(3):297--316, 2013.

\bibitem{neffke2013skill}
Frank Neffke and Martin Henning.
\newblock Skill relatedness and firm diversification.
\newblock {\em Strategic Management Journal}, 34(3):297--316, 2013.

\bibitem{neffke2011regions}
Frank Neffke, Martin Henning, and Ron Boschma.
\newblock How do regions diversify over time? {I}ndustry relatedness and the
  development of new growth paths in regions.
\newblock {\em Economic Geography}, 87(3):237--265, 2011.

\bibitem{neffke2011}
Frank Neffke, Martin Henning, Ron Boschma, Karl-Johan Lundquist, and Lars-Olof
  Olander.
\newblock The dynamics of agglomeration externalities along the life cycle of
  industries.
\newblock {\em Regional studies}, 45(1):49--65, 2011.

\bibitem{NelsonWinter1982}
Richard~R {Nelson} and Sidney~G {Winter}.
\newblock {\em An evolutionary theory of economic change}.
\newblock The Belknap Press of Harvard University Press, Cambridge, 1982.

\bibitem{newman2008physics}
Mark Newman.
\newblock The physics of networks.
\newblock {\em Physics today}, 61(11):33--38, 2008.

\bibitem{newman2006structure}
Mark~Ed Newman, Albert-L{\'a}szl{\'o}~Ed Barab{\'a}si, and Duncan~J Watts.
\newblock {\em The structure and dynamics of networks.}
\newblock Princeton university press, 2006.

\bibitem{newman2003structure}
Mark~EJ Newman.
\newblock The structure and function of complex networks.
\newblock {\em SIAM review}, 45(2):167--256, 2003.

\bibitem{newman2006modularity}
Mark~EJ Newman.
\newblock Modularity and community structure in networks.
\newblock {\em Proceedings of the national academy of sciences},
  103(23):8577--8582, 2006.

\bibitem{newman2004finding}
Mark~EJ Newman and Michelle Girvan.
\newblock Finding and evaluating community structure in networks.
\newblock {\em Physical review E}, 69(2):026113, 2004.

\bibitem{newman2020improved}
MEJ Newman, George~T Cantwell, and Jean-Gabriel Young.
\newblock Improved mutual information measure for clustering, classification,
  and community detection.
\newblock {\em Physical Review E}, 101(4):042304, 2020.

\bibitem{ocleryIreland}
Neave O'Clery, Eoin Flaherty, and Stephen Kinsella.
\newblock The network structure of knowledge flow.
\newblock {\em Forthcoming}, 2019.

\bibitem{onnela2012taxonomies}
Jukka-Pekka Onnela, Daniel~J Fenn, Stephen Reid, Mason~A Porter, Peter~J Mucha,
  Mark~D Fricker, and Nick~S Jones.
\newblock Taxonomies of networks from community structure.
\newblock {\em Physical Review E}, 86(3):036104, 2012.

\bibitem{orman2012comparative}
G{\"u}nce~Keziban Orman, Vincent Labatut, and Hocine Cherifi.
\newblock Comparative evaluation of community detection algorithms: a
  topological approach.
\newblock {\em Journal of Statistical Mechanics: Theory and Experiment},
  2012(08):P08001, 2012.

\bibitem{oclery2016}
N~O’Clery, A~Gomez, and Lora E.
\newblock The path to labour formality: urban agglomeration and the emergence
  of complex industries.
\newblock {\em {Harvard Center for International Development Working Paper}},
  2016.

\bibitem{o2019commuting}
Neave O’Clery, Rafael~Prieto Curiel, and Eduardo Lora.
\newblock Commuting times and the mobilisation of skills in emergent cities.
\newblock {\em Applied Network Science}, 4(1):118, 2019.

\bibitem{o2019modular}
Neave O’Clery, Eoin Flaherty, and Stephen Kinsella.
\newblock Modular structure in labour networks reveals skill basins.
\newblock {\em arXiv preprint arXiv:1909.03379}, 2019.

\bibitem{o2013observability}
Neave O’Clery, Ye~Yuan, Guy-Bart Stan, and Mauricio Barahona.
\newblock Observability and coarse graining of consensus dynamics through the
  external equitable partition.
\newblock {\em Physical Review E}, 88(4):042805, 2013.

\bibitem{palla2005uncovering}
Gergely Palla, Imre Der{\'e}nyi, Ill{\'e}s Farkas, and Tam{\'a}s Vicsek.
\newblock Uncovering the overlapping community structure of complex networks in
  nature and society.
\newblock {\em nature}, 435(7043):814--818, 2005.

\bibitem{peel2017ground}
Leto Peel, Daniel~B Larremore, and Aaron Clauset.
\newblock The ground truth about metadata and community detection in networks.
\newblock {\em Science advances}, 3(5):e1602548, 2017.

\bibitem{robins2007recent}
Garry Robins, Tom Snijders, Peng Wang, Mark Handcock, and Philippa Pattison.
\newblock Recent developments in exponential random graph (p*) models for
  social networks.
\newblock {\em Social networks}, 29(2):192--215, 2007.

\bibitem{Rosvall2009}
M.~Rosvall, D.~Axelsson, and C.T. Bergstrom.
\newblock The map equation.
\newblock {\em {The European Physical Journal Special Topics}}, 178(1):13--23,
  2009.

\bibitem{rosvall2009map}
Martin Rosvall, Daniel Axelsson, and Carl~T Bergstrom.
\newblock The map equation.
\newblock {\em The European Physical Journal Special Topics}, 178(1):13--23,
  2009.

\bibitem{schaub2017many}
Michael~T Schaub, Jean-Charles Delvenne, Martin Rosvall, and Renaud Lambiotte.
\newblock The many facets of community detection in complex networks.
\newblock {\em Applied network science}, 2(1):1--13, 2017.

\bibitem{Schaub2012}
M.T. Schaub, J.C. Delvenne, S.N. Yaliraki, and M.~Barahona.
\newblock Markov dynamics as a zooming lens for multiscale community detection:
  non clique-like communities and the field-of-view limit.
\newblock {\em {PloS One}}, 7(2), 2012.

\bibitem{sood2005voter}
Vishal Sood and Sidney Redner.
\newblock Voter model on heterogeneous graphs.
\newblock {\em Physical review letters}, 94(17):178701, 2005.

\bibitem{strogatz2001exploring}
Steven~H Strogatz.
\newblock Exploring complex networks.
\newblock {\em nature}, 410(6825):268--276, 2001.

\end{thebibliography}

\section*{Declarations}
\subsection*{Author's contributions}
NOC, DS and ML designed the study and wrote the paper. DS and ML carried out the analysis.  

\subsection*{Acknowledgements}
We would like to thank Ioan Alexandru Puiu and Renaud Lambiotte for their insightful suggestions and comments.

\subsection*{Funding}

This project (NOC and DS) was funded by a Turing-HSBC-ONS Economic Data Science Award 'Network modelling of the UK's urban skill base'. ML received support from the Skye Foundation Trust and the Oppenheimer Memorial Trust.

\subsection*{Availability of Data and Materials}

The methodology to reproduce the synthetic networks is outlined in the Supplementary Materials. The SRNs for Germany and Sweden are hosted by Frank Neffke \cite{Neffke2013SkillRelatedness} http://www.frankneffke.com/relatedness.html, and the SRN for the Netherlands is hosted by Dario Diodato \cite{diodato2014resilience} http://dariodiodato.com. The SRNs for Ireland and the UK are available upon request.


\subsection*{Competing interests}
The authors declare that they have no competing interests.


\appendix
\numberwithin{equation}{section}
\makeatletter 
\newcommand{\section@cntformat}{Appendix \thesection:\ }

\section*{Inter-industry labour flow networks}

Here, we further elaborate on the comparison of the modular structure of the United Kingdom Skill-Relatedness Network (SRN) to the Irish, German, Dutch and Swedish SRNs using the BiDir distance. 

Recall, the BiDir distance requires the networks being compared to be node-aligned (\textit{i.e.} all the networks need to have the same node set). To ensure that this is the case, we first convert the UK network so that industries are defined according to the NACE 1 industry classification. We then take the node intercept of all the networks being compared. Next, we elaborate on each of these steps.

Although the UK SRN has been constructed using the methodology of Neffke \emph{et al} \citep{Neffke2013SkillRelatedness}, due to data availability it was constructed using the 4-digit NACE 2 industry classification. Using the official concordance table provided by Eurostat, we convert the industries from a 4-digit NACE 2 to a 4-digit NACE 1 industry classification. We adopt the network-based convergence methodology of Diodato \cite{diodato2018network}. 

Once all the networks are in the same industry classification, we take the node intercept of all the networks being compared. We observe that the UK’s SRN has fewer industries compared to the other four countries. It is also much more sparse. This is because it was constructed from a longitudinal survey including just 1 per cent of UK workers (ASHE). On the other hand, all other networks were constructed using administrative data containing all workers within the formal sector. Hence, when comparing the UK SRN to those of other countries, only the subgraph induced by the set of overlapping nodes (and their corresponding edges) are compared in each case. 

\begin{figure}[!t]
    \centering
    \includegraphics[width=1\linewidth]{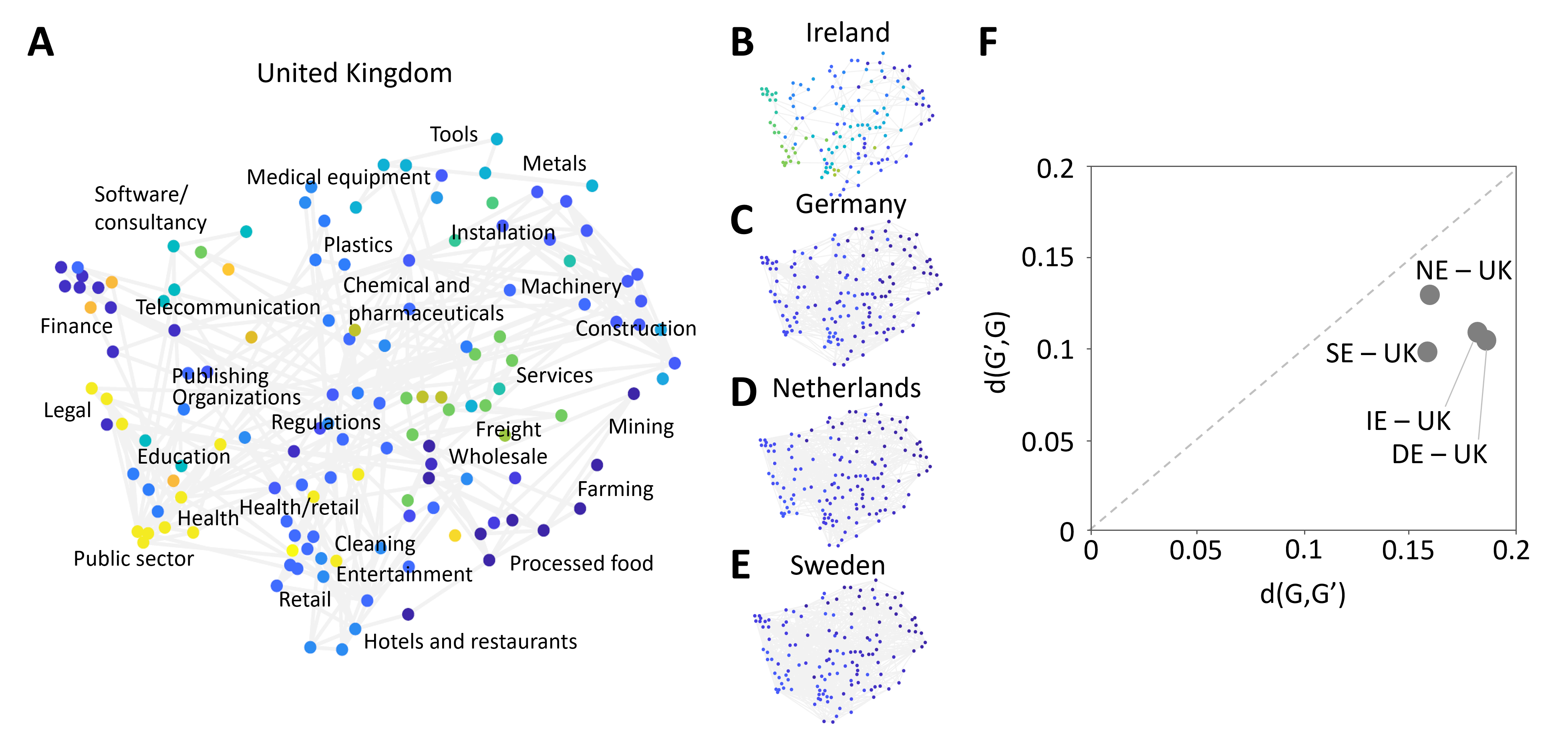}
    \caption{The pairwise comparison of the modular structure of the United Kingdom (UK) to four different countries' SRNs, namely: Ireland (IE), Germany (DE), the Netherlands (NE), and Sweden (SE). In (A) we show the UK SRN. The nodes represent industries and edges the skill-relatedness between the two corresponding industries. Nodes are coloured according to their community (found using the modularity optimization algorithm). The same node layout is used as in Figure~\ref{SRNetCom}. Labelling is added to indicate the general position of sectors within the network. Using the same layout we also show communities detected for IE, DE, the NE and SE in sub-figure (B) to (E). In (F) we show the resulting BiDir distance from the pairwise comparison of the UK SRN to the other countries' SRNs.}
    \label{SRNetCom2}
\end{figure}

In Figure~\ref{SRNetCom2} (A) We show a visualisation of the UK SRN. Each node in the network represents an industry and each edge the skill-relatedness between its two corresponding industries. The same node layout and node labelling is used as in Figure~\ref{SRNetCom} (A). The modular structure of the network is shown via the node colouring. The communities were detected using modularity \cite{newman2006modularity}. Specifically, the partition maximizing the modularity function from $10 000$ iterations of the Louvain algorithm is adopted. Using the same layout, we visualise the communities found for Irish, Germany, the Netherlands and Sweden on the right. Note that for all the networks only the induced sub-graph of the SRNs are shown.  

We now quantify the distance between the UK SRN and the other four countries' SRN using our BiDir distance. The pairwise comparison is shown in Figure \ref{SRNetCom2} (F). We observe that the modular structure of the UK network is very different to those of the other countries. We also observe that all the distances show an asymmetric relationship. For all four countries a smaller distance is obtained when projecting the community structure onto the UK network than vice versa. As the UK network is particularly sparse, the modularity algorithm unveils a more fine-grained community structure. Our measure is therefore picking up this difference in the scale at which the communities are most robust.

\section*{Technical details regarding SBM construction}

In Figure \ref{SBM} we can see the performance of our framework in three different families of SBMs. We now summarize how these three different families were generated. All networks consist of 120 nodes and are generated according to the following parameters: 
\begin{itemize}
	\item All networks in family $A$ consist of 4 identical blocks of size 30, and the SBM parameters are $P_{ij} = 0.20$ and $P_{ii} = 0.80, 0.60, 0.45, 0.35, 0.20$ for each successive network. Since for the last network, $A5$, both probabilities are the same, we do not have any a priori community structure, and we can use this network as a point of comparison.
    \item For all networks in family $B$, $P_{ij} = 0.20$ and $P_{ii} = 0.80$. But each successive networks splits some of the blocks of the previous one. So $B2$ split one of the original blocks into two, $B3$ splits the other original block to obtain 4 identical medium blocks. $B4$ splits one of these 4 blocks into 3 smaller ones, and $B5$ repeats that procedure on the other 3 medium blocks.
    \item $C1$ and $C2$ are produced by again using $P_{ij} = 0.20$ and $P_{ii} = 0.80$, with the first one split into 3 blocks and the second one into 4 blocks. $C3$ has the same blocks as $C2$, but  $P_{2,3} =P_{3,2} = 0.40$.
 \end{itemize}
 
 Furthermore, in Figure \ref{Res} we illustrate the use of the BiDir distance using a multi-resolution community detection algorithm. Here, we provide more detail regarding how the three nested SBMs were constructed. All three of the networks consists of 300 nodes and have implanted communities at $3$-levels. As illustrated in the adjacency matrices of each network, each network contains regions with one of 4 different shades of grey. These regions correspond to 4 different values of $P_{ij}$. From darkest to lightest the regions correspond to the following four values of $P_{ij} \in \{0.8, 0.53, 0.27, 0.15\}$. The size of the different regions for each network are given as follows: 
\begin{itemize}
 \item Network A consists of 12 identical blocks of size 25. Two of these blocks are then merged together to form 6 identical blocks of size 50. Similarly, two of these blocks are then grouped together to form 3 identical blocks of size 100.
 \item Network B, consists of 18 blocks of size $20,10,10,25,10,25, 20,10,10,25,10,25,\\ 20,10,10,25,10,25$, respectively. Two consecutive blocks are then joined together to create 9 blocks of size $30, 35, 35, 30, 35, 35, 30, 35, 35$, respectively. Finally, three consecutive blocks are joined together to create three identical blocks of size 100 - similar to Network A. 
 \item Network C, consists of 12 identical blocks of size 25, similar to Network A. These blocks are merged together to form four communities containing three blocks of size 75. Finally, two 75-sized blocks are joined together to form two communities of size 150.  
\end{itemize}

\end{document}